\documentclass[aps]{revtex4}%
\usepackage{amsfonts}
\usepackage{amsmath}
\usepackage{amssymb}
\usepackage{epsfig}
\usepackage{graphicx}%
\begin{document}
\title{Application of the Optimized Baxter Model to the hard-core attractive Yukawa system}
\author{Peter Prinsen$^{a}$*, Josep C. P\`amies$^{b}$, Theo Odijk$^{a}$ and Daan Frenkel$^{b}$}
\affiliation{$^{a}$Complex Fluids Theory, Faculty of Applied
Sciences, Delft University of Technology, Julianalaan 67, 2628 BC
Delft, The Netherlands\\ $^{b}$FOM Institute for Atomic and
Molecular Physics, Kruislaan 407, 1098 SJ Amsterdam, The
Netherlands}

\begin{abstract}

We perform Monte Carlo simulations on the hard-core attractive
Yukawa system to test the Optimized Baxter Model that was
introduced in [P.Prinsen and T. Odijk,  J. Chem. Phys.
\textbf{121}, 6525 (2004)] to study a fluid phase of spherical
particles interacting through a short-range pair potential. We
compare the chemical potentials and pressures from the simulations
with analytical predictions from the Optimized Baxter Model. We
show that the model is accurate to within 10 percent over a range
of volume fractions from 0.1 to 0.4, interaction strengths up to
three times the thermal energy and interaction ranges from 6 to 20
\% of the particle diameter, and performs even better in most
cases. We furthermore establish the consistency of the model by
showing that the thermodynamic properties of the Yukawa fluid
computed via simulations may be understood on the basis of one
similarity variable, the stickiness parameter defined within the
Optimized Baxter Model. Finally we show that the Optimized Baxter
Model works significantly better than an often used, naive method
determining the stickiness parameter by equating the respective
second virial coefficients based on the attractive Yukawa and
Baxter potentials.

\vspace{10pt}

*Corresponding author: p.prinsen@tnw.tudelft.nl

\end{abstract}
\maketitle

\section*{I. INTRODUCTION}

In a recent paper \cite{PRI} two of us devised a method to
approximate systematically a system of spherical hard particles
that interact through a short-range pair potential by a system of
particles interacting via an effective Baxter potential
\cite{BAX}. The latter consists of a hard-core repulsion and a
sticky attraction at the surface of the particles which is
computed by a variational principle for the free energy. The
original potential was a sum of attractive and repulsive
contributions (i.e. a square well plus a Debye-H\"uckel
interaction \cite{PRI}) but the variational method also applies to
a purely attractive interaction provided its range is sufficiently
smaller than the particle diameter. The advantage of approximating
the original interaction by the Baxter potential is that the fluid
phase of the Baxter model has been studied extensively, both
theoretically \cite{BAX,WAT,BA1,CUM,BA2,POS,STE,GAZ} and in
computer simulations \cite{SEA,KRA,MI1,MI2}. This means that once
the correspondence between the two systems has been established,
all the results of the Baxter model can be used for the original
system.

In the Optimized Baxter Model (OBM) \cite{PRI}, the free energy of
the actual system is functionally expanded in terms of the Mayer
function where the reference state is a suspension of hard spheres
interacting via an effective sticky potential. The stickiness
parameter associated with the latter is determined by setting the
first-order term in this expansion equal to zero. This constitutes
a variational principle since the second-order term turns out to
be either positive or negative definite. Nevertheless, the exact
nature of the expansion is difficult to assess analytically. For
instance, there may be mathematical problems arising from the
limiting procedure in which the range of the effective adhesion
goes to zero as its magnitude becomes infinitely large \cite{PRI}.
Thus, a computational test of the OBM is important.

The model system we consider consists of hard-sphere particles
with an attractive Yukawa interaction
\begin{equation}
\frac{U_{Y}(x)}{k_{B}T}=\left\{
\begin{array}
[c]{ll}
\infty & \hspace{15pt} 0\leq x<2\\
-\beta\frac{e^{-\kappa(x/2-1)}}{x/2} & \hspace{15pt} x>2
\end{array}
\right.\label{yuk}
\end{equation}
\[
x\equiv\frac{r}{a}.
\]
Here $a$ is the radius of the particles, $k_{B}$ is Boltzmann's
constant, $T$ is the temperature, $\beta$ is the dimensionless
well depth and $a/\kappa$ is a measure of the range of the
attractive tail (if we set the actual well depth $\equiv$ unity,
$\beta$ may be viewed as identical with $1/k_{B}T$). The somewhat
awkward factors $1/2$ in front of the dimensionless coordinate $x$
are caused by the fact that we want to scale all distances by $a$
as in Ref. \cite{PRI} whereas, at the same time, we wish to use
the same notation as in other papers on simulation studies of the
same system \cite{HAG,DIJ} where distances are scaled by the
diameter $2a$ of the particles.

The liquid-solid coexistence of this system has been studied
before at various values of $\kappa$ \cite{HAG,DIJ}. These papers
do not report the chemical potentials and pressures at coexistence
however. We need these data to compare them to the predictions of
the OBM to test its validity. We therefore perform new simulations
to determine the volume fraction, chemical potential and pressure
at various points along the phase boundaries. Moreover, we also
determine the chemical potential and pressure within the fluid
region of the phase diagram so as to gauge the accuracy of the OBM
at lower concentrations..

We start by reviewing equations relevant to the OBM as applied to
the Yukawa potential (Eq. (\ref{yuk})) in the next section. In
section III we describe the numerical simulations which, in
section IV, are compared with the theoretical predictions.

\section*{II. THEORY}

Here we give a short summary of the theory developed in Ref.
\cite{PRI}. The relevant equations needed to determine the
effective adhesion parameter $\tau$ and some of the thermodynamic
properties of the system are presented here. For details of the
derivation we refer to Ref. \cite{PRI} and references mentioned
there.

We consider a system of spherical particles of radius $a$. The
interaction $U$ between the particles is pairwise additive and
consists of a hard sphere repulsion plus a short-range interaction
$U_{1}$ that is either purely attractive or consists of a
combination of attractive and repulsive interactions (range $\ll
a$). In the latter case the attraction has to be strong enough to
compensate for the repulsion---we come back to this issue later.
For convenience, all distances are scaled by the radius $a$ of the
particles so we have
\begin{equation}
U(x)=\left\{
\begin{array}
[c]{ll}%
\infty & \hspace{15pt} 0\leq x<2\\
U_{1}(x) & \hspace{15pt} x>2
\end{array}
\right.
\end{equation}
\[
x\equiv\frac{r}{a}.
\]
We wish to replace this system by a suspension of adhesive hard
spheres with the same radius which is our reference state. The
interaction of the latter is given by the adhesive hard sphere
(AHS) potential of Baxter \cite{BAX}
\begin{equation}
\frac{U_{AHS}(x)}{k_{B}T}=\left\{
\begin{array}
[c]{ll}%
\infty & \hspace{15pt} 0\leq x<2\\
\ln\frac{12\tau\omega}{2+\omega} & \hspace{15pt} 2\leq x\leq2+\omega\\
0 & \hspace{15pt} x>2+\omega
\end{array}.
\right.\label{ahs}
\end{equation}
Here, $\tau$ is a positive constant that we wish to determine and
which signifies the strength of the effective adhesion and the
limit $\omega\downarrow0$ has to be taken after formal
integrations. The reason for approximating the original system by
the AHS system is that the latter has been conveniently solved in
the Percus-Yevick approximation \cite{BAX}. This means that once
the correspondence between the two systems has been established by
appropriately choosing $\tau$, other properties like for example
the chemical potential, the pressure and the compressibility of
the system can easily be computed analytically from the solution
of the AHS system.

We first describe how to choose the stickiness parameter $\tau$.
In the limit of vanishing density this is done by equating second
virial coefficients since we must equate the respective free
energies of the two systems.
\begin{equation}
B_{2}=2\pi
a^{3}\int_{0}^{\infty}dx\;x^{2}\left(1-e^{-U\left(x\right)/k_{B}T
}\right)\equiv
B_{2}^{AHS}=B_{2}^{HS}\left(1-\frac{1}{4\tau_{0}}\right).
\end{equation}
This amounts to choosing
\begin{equation}
\tau_{0}=\frac{2}{3\int_{2}^{\infty}dx\;x^{2}\left(e^{-U_{1}
\left(x\right)/k_{B}T}-1\right)}.\label{B2}
\end{equation}
Here, $B_{2}^{HS}=16\pi a^{3}/3$ is the second virial coefficient
of a solution of hard spheres. At finite densities this procedure
necessarily breaks down, however, because the higher virials come
into play. The stickiness parameter $\tau$, which depends on the
density, has to be obtained by identifying the free energy of the
actual system with that of the reference state as well as
possible. In the functional expansion of the excess free energy in
terms of the Mayer function \cite{AND}, we then demand that the
first order correction vanishes. This leads to the condition
\cite{PRI}
\begin{equation}
\int_{0}^{\infty}dx\;x^{2}(e^{-U(x)/k_{B}T}-1)\widetilde{g}(x)=
\frac{2\lambda}{3}, \label{dett}
\end{equation}
where $\widetilde{g}(x)$ is the regular part of the pair
correlation function $g(x)$ of the reference AHS system and
\begin{equation}
\lambda=\frac{6(\eta+(1-\eta)\tau)}{\eta(1-\eta)}
\left(1-\sqrt{1-\frac{\eta(2+\eta)}{6(\eta+(1-\eta)\tau)^2}}\right)
\label{lambda}
\end{equation}
with $\eta$ the volume fraction of particles. For $x\,<2$,
$\widetilde{g}(x)$ equals zero owing to the hard-core repulsion,
whereas $\widetilde{g}(x)$ tends to unity for large $x$. Since the
interaction $U(x)$ is of short range, we approximate
$\widetilde{g}(x)$ in the interval $2\leq x\leq4$ by the first two
terms of its Taylor expansion \cite{PRI}
\begin{equation}
\widetilde{g}(x)\simeq\left\{
\begin{array}
[c]{ll}
0 & \hspace{15pt} x<2\\
G(1+H(x-2)) & \hspace{15pt} 2\leq x\leq4\\
1 & \hspace{15pt} x>4
\end{array}.\label{distr}
\right.
\end{equation}
Here, we define
\begin{equation}
G=\lambda\tau\label{G}
\end{equation}
and
\begin{equation}
H=\frac{\eta}{2\tau(1-\eta)}\left(
\frac{\eta(1-\eta)}{12}\lambda^{2}
-\frac{1+11\eta}{12}\lambda+\frac{1+5\eta}{1-\eta}-\frac{9(1+\eta)}
{2(1-\eta)^{2}}\frac{1}{\lambda}\right).\label{H}
\end{equation}
At a given volume fraction $\eta$, $\tau$ can then be determined
iteratively from Eqs. (\ref{dett})-(\ref{H}). An iteration scheme,
which converges fast, consists of choosing a starting value of
$\tau$, determining $\lambda$ from Eq. (\ref{lambda}), then $\tau$
from
\begin{equation}
\tau=\frac{2\lambda-3\int_{4}^{\infty}dx\;x^{2}(e^{-U(x)/k_{B}T}-1)}
{3\lambda\int_{2}^{4}dx\;x^{2}(1+H(x-2))(e^{-U(x)/k_{B}T}-1)},\label{itt}
\end{equation}
$\lambda$ again from Eq. (\ref{lambda}) and so on until convergence
to the required accuracy is achieved.

There are two cases in which the above method does not yield
meaningful results. The first occurs when the short-range
interaction has both attractive and repulsive components in the
event that the repulsion is too strong in comparison with the
attraction. The total interaction is then effectively repulsive in
nature so it is clear that a suspension of particles interacting
in such a way cannot be approximated by an AHS system. In this
case, the iteration scheme described above leads to a $\tau$ which
keeps on increasing and does not converge. If $\tau_{0}$ is
negative in the limit of vanishing density (Eq. (\ref{B2}))
implying a net repulsion, it is advisable not to compute $\tau$
also, even though it could lead to positive values at higher
densities. Secondly, the attraction may be too strong. There
exists a critical value of the stickiness parameter $\tau_{c}$
below which there is a range of densities for which there is no
real solution of $\lambda$
\begin{equation}
\tau_{c}=\frac{2-\sqrt{2}}{6}.
\end{equation}
This means that if the attraction is strong enough (i.e. when
$\tau$ is too small), there will not be a positive real solution
to Eq. (\ref{itt}). In this case the iteration scheme would
produce complex values of $\tau$.

We now state several thermodynamic properties resulting from the
solution of the Baxter model which we will need further on. To
compute the pressure $P$ and the chemical potential $\mu$ we use
the expressions derived via the compressibility route
\cite{BAX,BA1}
\begin{equation}
\frac{Pv_{0}}{k_{B}T}=\frac{\eta(1+\eta+\eta^{2})}{(1-\eta)^{3}}-
\frac{\eta^{2}(1+\eta/2)}{(1-\eta)^2}\lambda+
\frac{\eta^{3}}{36}\lambda^{3} \label{pressure}
\end{equation}
and
\begin{equation}
\frac{\mu-\mu_{0}}{k_{B}T}=\ln\frac{\eta}{1-\eta}+
\frac{3\eta(4-\eta)}{2(1-\eta)^2}+\frac{Pv_{0}}{k_{B}T}+J.
\label{chempot}
\end{equation}
Here $v_{0}=4\pi a^{3}/3$ is the volume of a particle,
\begin{equation}
J=\frac{3}{2}\eta^{2}\lambda^{2}-
\frac{3\eta(1+4\eta)}{(1-\eta)}\lambda
+\frac{6\eta(2+\eta)}{(1-\eta)^{2}}
-\frac{18\eta}{1-\eta}\tau-\frac{6(\tau-\tau_{c})^{2}}{\tau_{c}(1-6\tau_{c})}
\ln\left|\frac{\lambda(1-\eta)-\tau_{c}^{-1}}{\tau^{-1}-\tau_{c}^{-1}}\right|
+\frac{6\tau_{c}(18\tau\tau_{c}-1)^{2}}{1-6\tau_{c}}
\ln\left|\frac{\lambda(1-\eta)-18\tau_{c}}{\tau^{-1}-18\tau_{c}}\right|
\end{equation}
is the contribution to the chemical potential that vanishes in the
hard-sphere limit ($\tau\rightarrow\infty$) and the chemical
potential of the reference state (in the context of the Baxter
model) is given by
\begin{equation}
\frac{\mu_{0}}{k_{B}T}= \ln \frac{1}{v_{0}}\left(\frac{h^{2}}{2\pi
mk_{B}T}\right)^{3/2}
\end{equation}
where $h$ is Planck's constant and $m$ is the mass of a particle.

\section*{III. SIMULATIONS}

We perform Monte Carlo simulations at constant volume $V$ and
temperature $T$ on a system of $N=256$ hard spheres with a
short-range Yukawa attraction so we have (compare with Eq.
(\ref{yuk}))
\begin{equation}
\frac{U_{1}(x)}{k_{B}T}=-\beta\frac{e^{-\kappa(x/2-1)}}{x/2}.
\end{equation}
We introduce a cutoff at $x=4$, so that $U(x)=0$ for $x>4$. We
determine the Helmholtz free energy per particle $f_{N}$ at a
chosen set of parameters of $\beta$, $\kappa$ and $\eta$ by
thermodynamic integration at constant $\kappa$ and $\eta$,
starting from the known free energy per particle of the hard
sphere system ($\beta=0$) which is defined at the same volume
fraction (see e.g. \cite{FRE})
\begin{equation}
\frac{f_{N}(\eta,\beta)-f_{N}(\eta,0)}{k_{B}T}=\int_{0}^{\beta}d\beta'\;\frac{1}{\beta'}\left\langle
\frac{U_{1}}{k_{B}T}\right\rangle_{N}\label{int}.
\end{equation}
Here $\langle U_{1}\rangle_{N}$ is the average energy per particle
where the average is computed in the state with $\beta=\beta'$.
From this we determine the equation of state $z_{N}(\eta,\beta)$
\begin{equation}
z_{N}(\eta,\beta)-z_{N}(\eta,0)=\frac{P_{N}(\eta,\beta)}{\rho
k_{B}T}-\frac{P_{N}(\eta,0)}{\rho k_{B}T}=
\eta\frac{\partial}{\partial\eta}
\left[\frac{f_{N}(\eta,\beta)-f_{N}(\eta,0)}{k_{B}T}\right],\label{eos}
\end{equation}
where the particle density $\rho$ is related to the volume
fraction by $\rho v_{0}=\eta$, and the chemical potential
$\mu_{N}(\eta,\beta)$ is given by
\begin{equation}
\frac{\mu_{N}(\eta,\beta)}{k_{B}T}-\frac{\mu_{N}(\eta,0)}{k_{B}T}\simeq
\frac{\partial}{\partial\eta}\eta\left[\frac{f_{N}(\eta,\beta)-f_{N}(\eta,0)}{k_{B}T}\right].
\label{chempot2}
\end{equation}
The expression for the pressure is exact for a system consisting
of a finite number of particles $N$, whereas that for the chemical
potential has an error of order $N^{-1}$, because in the
simulations we change only the volume $V$ of the box leaving the
number of particles invariant. (See the appendix for details). For
the equation of state of the pure hard-sphere system we have
\begin{equation}
z^{l}(\eta,0)=1+\frac{4\eta+1.216224\eta^2+1.246720\eta^3}
{1-2.195944\eta+1.210352\eta^2}\label{sp1}
\end{equation}
valid when the system is a fluid \cite{SP1}. It is quadrature to
determine the chemical potential
\begin{equation}
\frac{\mu^{l}(\eta,0)}{k_{B}T}=\ln\eta-1+z^{l}(\eta,0)+
\int_{0}^{\eta}d\eta'\frac{z^{l}(\eta',0)-1}{\eta'}.
\end{equation}
For the pressure of the hard sphere (fcc) solid we use \cite{SP2}
\begin{equation}
z^{s}(\eta,0)=\frac{3}{1-\frac{6}{\pi\sqrt{2}}\eta}
-\frac{0.5921\left(\frac{6}{\pi\sqrt{2}}\eta-0.7072\right)}
{\frac{6}{\pi\sqrt{2}}\eta-0.601}
\end{equation}
and for the chemical potential
\begin{equation}
\frac{\mu^{s}(\eta,0)}{k_{B}T}=\ln\eta-1+z^{s}(\eta,0)+\frac{f^{s}(\eta_0)}{k_{B}T}+
\int_{\eta_{0}}^{\eta}d\eta'\frac{z^{s}(\eta',0)-1}{\eta'},
\end{equation}
where $f^{s}(\eta_{0})/k_{B}T=5.91889(4)$ is the free energy of
the hard sphere solid in the thermodynamic limit
$N\rightarrow\infty$ at volume fraction $\eta_{0}=0.5450$
\cite{POL}. We thus calculate the pressure and the chemical
potential of the system in the thermodynamic limit using
\begin{equation}
\frac{P(\eta,\beta)v_{0}}{k_{B}T}\simeq\eta z^{i}(\eta,0)+
\eta^{2}\frac{\partial}{\partial\eta}\int_{0}^{\beta}d\beta'\;\frac{1}{\beta'}\left\langle
\frac{U_{1}}{k_{B}T}\right\rangle_{N},\label{pres}
\end{equation}
and
\begin{equation}
\frac{\mu(\eta,\beta)}{k_{B}T}\simeq\frac{\mu^{i}(\eta,0)}{k_{B}T}+
\frac{\partial}{\partial\eta}\eta\int_{0}^{\beta}d\beta'\;\frac{1}{\beta'}\left\langle
\frac{U_{1}}{k_{B}T}\right\rangle_{N}\label{chem}
\end{equation}
where $i=s,l$. These expressions are not exact but correct to
order $N^{-1}$ because the number of particles in the simulations
is finite (see the appendix for details).

To determine the average energy per particle $\langle
U_{1}/k_{B}T\rangle_{N}$ we need to initiate the simulation by
choosing a convenient starting configuration. In the case that the
system is a solid, we assume it is an fcc crystal at the
appropriate density. For the liquid, a configuration at the
required density is initiated by putting the particles in the box
at random and then running the simulation until the particles no
longer overlap. This is done at a low value of $\beta$ and we then
use this starting configuration for all values of $\beta$ at the
same density. The simulation is then run for 10,000 cycles (i.e.
trial moves per particle) at the relevant value of $\beta$ to
determine the appropriate maximum displacement of a particle at an
acceptance probability of a particle displacement of 0.40. The
maximum displacement is then fixed and the simulation is run for
another 10,000 cycles for the system to equilibrate. Finally, the
average energy per particle is measured every 100 cycles during
another 50,000 cycles.

To perform the integration in Eq. (\ref{int}) we run simulations
at values of $\beta$ ranging from $0.1$ until the appropriate
value at intervals of 0.1. A simulation at $\beta=0.02$ is also
performed. We then fit the points to a curve and use this to
perform the integration. To determine the density dependence of
the free energy $f_{N}$ about a certain density, we compute the
free energy at about 10 values of the density close to it, at
intervals of 0.1. We again fit these to a curve which is used in
Eqs. (\ref{pres}) and (\ref{chem}) to determine the chemical
potential and the pressure at the desired density.

\section*{IV. RESULTS AND DISCUSSION}

\subsection*{A. Phase equilibrium}

We first test the Optimized Baxter Model on the fluid-solid
coexistence of hard spheres with Yukawa attraction. This
coexistence has been studied before via computer simulations
\cite{HAG,DIJ} but these papers did not report the pressure and
chemical potential, data we do need here.

At a given strength $\beta$ and inverse range $\kappa$ of the
attraction, we compute the volume fractions of the coexisting
fluid and solid phases by equating the pressures and the chemical
potentials in the respective phases. This is done for $\kappa=7$
and 9, and $\beta$ running from 0 to 2 at intervals of 0.25 (see
Table I). Our phase boundaries at $\kappa=7$ agree well with those
computed by Dijkstra \cite{DIJ} where the same method was used
though with a smaller system of $N=108$ particles. The deviation
in volume fraction is at most 2\% (we determined the phase
boundaries from a plot presented in Ref. \cite{DIJ}, so this may
account for part of the difference). At low values of $\beta$ the
agreement with the simulations of Hagen and Frenkel \cite{HAG} is
also good, but with increasing $\beta$ the difference between
their phase boundary on the fluid side and ours becomes
appreciable until our prediction of the volume fraction is about
20\% higher than theirs at $\beta=2$. We note that in Ref.
\cite{HAG} a different method was used to determine the phase
boundary. The phase boundaries on the solid side do agree within
3\%. We regain essentially the same picture at $\kappa=9$ though
the difference in the phase boundaries at the fluid side is less
pronounced (about 14\% at $\beta=2$). The phase diagram at
$\kappa=9$ was not determined in Ref. \cite{DIJ}.

Next, we use the OBM to determine the effective stickiness
parameter $\tau$ (Eq. (\ref{itt})) and the coexistence curve. By
way of comparison, we also evaluate $\tau_{0}$ by equating the
respective second virial coefficients of the attractive Yukawa
interaction and the Baxter potential (see Eq. (\ref{B2})) and
computing the properties of the resulting Baxter fluid. We will
refer to this as the $B_{2}$ method which is strictly correct only
at very low concentrations. We employ Eqs. (\ref{pressure}) and
(\ref{chempot}) to calculate the pressure and the chemical
potential from the volume fractions and the respective values of
$\tau$ from the two methods. These predictions are compared with
the simulations in Fig. 1 ($\kappa=7$) and Fig. 2 ($\kappa=9$). It
is clear from the figures that the predictions of the OBM are
significantly better than those via the $B_{2}$ method along the
whole phase boundary. The OBM is actually quite accurate to within
a few percent. Recall that at $\beta=0$, i.e. in the absence of
attraction, the two volume fractions predicted by the two methods
necessarily coincide simply because $\tau=\infty$ in that case.
However, they do not agree with the simulations which is due to
the fact that we use the accurate equation of state (Eq.
(\ref{sp1})) in the simulations whereas the analytical theory is
of course approximate. The latter overestimates the pressure and
the chemical potential in the case of hard spheres.

\subsection*{B. Consistency test}

We next check the consistency of the OBM. The Baxter model itself
has been solved in the Percus-Yevick approximation and we here use
the compressibility route to obtain the thermodynamic properties.
We know, however, that in the case of the hard sphere system, the
analytical calculations via the Percus-Yevick approximation and
the compressibility route are too high (e.g. at $\eta=0.4$, both
the pressure and the chemical potential are overestimated by 4\%).
We therefore seek to test the argumentation within the OBM in a
way which is less sensitive to approximations inherent in the
Baxter model. For instance, we note that the stickiness parameter
$\tau$ in the OBM merely depends on the properties of the
distribution function $\widetilde{g}$ very close to the sphere
(see Eq. (\ref{distr})). Though this does depend on the
Percus-Yevick approximation, it stands to reason that the
functions $G$ and $H$ are more robust than the oscillatory
behavior which $\widetilde{g}$ actually displays in full (and
which is implicit in Eqs. (\ref{pressure}) and (\ref{chempot})).
Thus, in the following simulations, we investigate whether
similarity is achieved with respect to the parameter $\tau$ as
given by Eq. (\ref{itt}).

Our procedure is as follows. We start at a given volume fraction.
Next we choose a set of values of the inverse range of the Yukawa
potential (i.e. $\kappa=5$, 7, 9, 11, 13 and 15). We then fix a
certain value of the stickiness parameter $\tau$ and compute the
concomitant value $\beta$ for each $\kappa$ with the help of Eq.
(\ref{itt}). If similarity does apply, the thermodynamic
properties should depend solely on $\tau$ and $\eta$ i.e. they
ought to be independent of $\kappa$ at constant $\tau$.

We have performed this test on simulations in a suitable range of
volume fractions $\eta$ and stickiness parameters $\tau$ with
associated interaction parameters $\kappa$ and $\beta$ as chosen
above. (See Figs. 3 to 6.) In some cases the attraction is so
strong in terms of $\beta$ that the simulated fluid is actually in
a metastable region. In effect, if the system were macroscopic,
phase separation into fluid and crystal phases would occur. (See
for example the relevant figures in Refs. \cite{HAG,DIJ}). This
happens when $\tau=0.1$, for example in the case $\kappa=15$ where
we have $\beta\approx 3.5$ for all four volume fractions and in
the case $\kappa=7$ where $\beta\approx 2.4$. Despite the
pre-emption of phase separation, we may still determine the
pressure and chemical potential as if the phases were stable.

We first note that the simulated thermodynamic properties are
generally quite independent of $\kappa$. (See the filled symbols
in Figs. 3 to 6). This implies that $\tau$ is indeed a useful
similarity variable and the OBM is a consistent approximation
scheme. The variation in the pressures and chemical potentials
computed by simulation is only a few percent with few exceptions
(e.g. in some instances, at lower values of $\kappa$ when the
volume fraction is 0.4, or when $\tau=0.1$ at the higher
densities; in the latter case, scrutiny of the simulation
snapshots shows that gelation seems to be occurring---note that
the attraction is so strong that we are now well beyond the
percolation threshold \cite{MI1}).

Next, it is of interest to compare the magnitudes of the simulated
thermodynamic properties with those computed with the help of the
OBM (see the curves in Figs. 3 to 6 which are horizontal because
$\tau$ was forced to be constant in each case). The analytical
predictions are virtually quantitative, except at those densities
at $\tau=0.1$ where gelation seems to occur as discussed above and
with regard to some of the pressures at higher concentrations. The
latter are overestimated at $\tau=0.5$ and 1 in Figs. 5 and 6
which we attribute to deficiencies in the Baxter model itself,
since the simulational data are quite independent of $\kappa$ as
stressed above.

For the sake of comparison we have also displayed thermodynamic
properties computed by the $B_{2}$ method. At a certain $\kappa$
and $\beta$ we evaluate $\tau_{0}$ with the help of Eq. (\ref{B2})
using the Yukawa interaction Eq. (\ref{yuk}) (thus $\tau_{0}$ is
not constant like $\tau$) and then calculate the pressure and
chemical potential within the Baxter model. The $B_{2}$ method
works well at low concentrations (see Fig. 3), which is not
surprising since neglecting to variationally adjust virials higher
than second is not so crucial in this case. However, the $B_{2}$
method worsens progressively as the concentration increases and
ultimately becomes unreliable (see Figs. 4-6). This is of course
expected: the $B_{2}$ method merely adjusts a single coefficient
$B_{2}$ whereas the free energy itself is variationally optimized
in the OBM.

We conclude that the Optimized Baxter Model is a convenient
quantitative, analytical theory for computing the thermodynamic
properties of a fluid of hard spheres interacting by an attraction
of short range. Moreover, the variational scheme used in deriving
the OBM is consistent, especially when the range of the potential
is short i.e. less than approximately 10 \% of the particle
diameter ($\kappa\gtrsim 10$). Overall, the OBM is accurate to
within 10 percent, except under conditions of very strong
attraction at high volume fractions ($\tau=0.1$, $\eta=0.3$ and
0.4), and it is actually much more precise in most cases.

\subsection*{Acknowledgements}

The work of the FOM institute is part of the research program of
the Foundation for Fundamental Research on Matter (FOM) and was
made possible through financial support by the Dutch Foundation
for Scientific Research (NWO).

\section*{V. APPENDIX A}

Here we show that the error incurred in Eq. (\ref{chempot2}) for
the chemical potential of the $N$-particle system is of order
$N^{-1}$, whereas Eq. (\ref{eos}) for the equation of state is
exact. We also prove that the error in the free energy of the
system is of order $N^{-1}$.

Our simulations are carried out at a constant number of particles
$N$. Hence, we modify the volume fraction $\eta$ by altering the
volume of the simulation box. The free energy difference per
particle
\begin{equation}
\Delta f_{N}(\eta,\beta)\equiv
f_{N}(\eta,\beta)-f_{N}(\eta,0)=k_{B}T
\int_{0}^{\beta}d\beta'\;\frac{1}{\beta'}\left\langle
\frac{U_{1}}{k_{B}T}\right\rangle_{N}
\end{equation}
is determined as a function of the volume fraction, so in effect
it is a function of $\eta$ (or $\rho$) and $N$ (and of course
$\beta$ and $\kappa$). The exact equation of state
$z_{N}(\eta,\beta)$ for the $N$-particle system is then
\begin{equation}
z_{N}(\eta,\beta)\equiv -\frac{1}{\rho k_{B}T}\left(\frac{\partial
F_{N}(\eta,\beta)}{\partial V}\right)_{N,T}=z_{N}(\eta,0)+
\eta\frac{\partial}{\partial\eta}\frac{\Delta
f_{N}(\eta,\beta)}{k_{B}T}, \label{eos2}
\end{equation}
where $F_{N}(\eta,\beta)=Nf_{N}(\eta,\beta)$ and the exact chemical
potential is
\begin{equation}
\frac{\mu_{N}(\eta,\beta)}{k_{B}T}\equiv \frac{1}{k_{B}T}
\left(\frac{\partial F_{N}(\eta,\beta)}{\partial
N}\right)_{V,T}=\frac{\mu_{N}(\eta,0)}{k_{B}T}+
\frac{\partial}{\partial\eta}\frac{\eta\Delta
f_{N}(\eta,\beta)}{k_{B}T}-\frac{1}{N}\frac{\partial}{\partial
N^{-1}}\frac{\Delta f_{N}(\eta,\beta)}{k_{B}T}.\label{chempot3}
\end{equation}
Here, and in the rest of the appendix, we have switched to the new
independent variables $\eta$ and $N$ so that derivatives with
respect to $\eta$ are taken at constant $N$ and derivatives with
respect to $N$ are taken at constant $\eta$. We see from Eqs.
(\ref{eos2}) and (\ref{chempot3}) that Eq. (\ref{chempot2}) has an
error of order $N^{-1}$ whereas Eq. (\ref{eos}) is exact.

We now assume that we may Taylor expand $\Delta f_{N}(\eta,\beta)$
for small values of $N^{-1}$ at constant volume fraction. It's not
obvious that this is allowed. In the case of a crystal for
example, the first-order correction to the free energy per
particle due to the fact that the number of particles is finite,
is of order $N^{-1}\ln N$ \cite{HOO,POL}. This correction is the
same for systems of identical numbers of particles however,
regardless of the interaction. Since our $f_{N}$ is the difference
in the free energies per particle pertaining to the two respective
crystals (with different pair potentials), the $O(N^{-1}\ln N)$
corrections simply cancel. Moreover, from Ref. \cite{POL} we know
that the leading higher order corrections to the free energy per
particles are of order $N^{-1}$. These deliberations are confirmed
in Figs. 7 and 8 which show that the leading corrections to the
average dimensionless energy per particle $\langle
U_{1}/k_{B}T\rangle_{N}$ are indeed of order $N^{-1}$ at the
representative values $\beta=1$, $\kappa=15$ and $\rho
(2a)^{3}=0.4$ ($\eta=\pi/15\approx 0.20944$) for the fluid and
$\rho (2a)^{3}=1.2$ ($\eta=\pi/5\approx 0.62832$) for the solid.
Therefore, we conclude that the free energy per particle in a
system containing an infinite number of particles is given by
\begin{equation}
\Delta f_{\infty}(\eta,\beta)=\Delta
f_{N}(\eta,\beta)+O\left(\frac{1}{N}\right).
\end{equation}
In the same manner, the equation of state $z_{\infty}(\eta,\beta)$
is then
\begin{equation}
z_{\infty}(\eta,\beta)=z_{\infty}(\eta,0)+
\eta\frac{\partial}{\partial\eta}\frac{\Delta
f_{N}(\eta,\beta)}{k_{B}T}+O\left(\frac{1}{N}\right),
\end{equation}
and the chemical potential is expressed by
\begin{equation}
\frac{\mu_{\infty}(\eta,\beta)}{k_{B}T}=\frac{\mu_{\infty}(\eta,0)}{k_{B}T}+
\frac{\partial}{\partial\eta}\frac{\eta\Delta
f_{N}(\eta,\beta)}{k_{B}T}+O\left(\frac{1}{N}\right).
\end{equation}

\section*{VI. APPENDIX B}

We estimate the second-order correction to the free energy (see
Appendix C of Ref. \cite{PRI})
\begin{equation}
\Delta=\frac{9}{4}\eta^{2}Y.\label{Delta}
\end{equation}
This correction leads for instance to a correction to the
dimensionless pressure $P v_{0}/k_{B}T$ approximately equal to
$-2\eta\Delta$ \cite{PRI}. The first part of the analysis in
Appendix C of Ref. \cite{PRI} is also useful here and we again
approximate $Y$ by
\begin{equation}
Y\simeq\frac{2}{3}\left(9G+10GH-12+\frac{\lambda}{2}\right) \left[
\int_{2}^{\infty}dt\;tB(t)\right]^{2}+\left[ \int_{2}^{\infty
}dt\;tB(t)\right]  \left[
\int_{2}^{\infty}ds\;s^{3}B(s)\right],\label{Y}
\end{equation}
where
\begin{equation}
B(x)\equiv g(x)\left[\exp\left(-\frac{U_{Y}(x)}{k_{B}T}+
\frac{U_{AHS}(x)}{k_{B}T}\right)-1\right] \label{Bx}
\end{equation}
and $\lambda$, $G$ and $H$ are given by Eqs.
(\ref{lambda})-(\ref{H}). We split the pair distribution function
$g(x)$ in the reference state into $g_{\omega}(x)$ and a regular
part $\widetilde{g}(x)$ given by Eq. (\ref{distr}) (see also Ref.
\cite{PRI}):
\begin{equation}
g(x)=\widetilde{g}(x)+g_{\omega}(x)
\end{equation}
with
\begin{equation}
g_{\omega}(x)=\left\{
\begin{array}
[c]{ll}
0 & x<2\\
\frac{\lambda(2+\omega)}{12\omega}+O(1)\hspace{15pt} & 2\leq x\leq2+\omega\\
0 & x>2+\omega.
\end{array}
\right.
\end{equation}
We then insert the expressions for the potentials Eqs. (\ref{yuk})
and (\ref{ahs}) into Eq. (\ref{Bx}) and derive in the limit
$\omega\rightarrow 0$
\begin{equation}
\int_{2}^{\infty}\mbox{d}x\;xB(x)=-\frac{\lambda}{3}+
G\int_{2}^{\infty}\mbox{d}x\;x(1+H(x-2))(e^{-U_{Y}(x)/k_{B}T}-1)
+O(e^{-\kappa})\label{x1}
\end{equation}
and
\begin{equation}
\int_{2}^{\infty}\mbox{d}x\;x^{3}B(x)=-\frac{4\lambda}{3}+
G\int_{2}^{\infty}\mbox{d}x\;x^{3}(1+H(x-2))(e^{-U_{Y}(x)/k_{B}T}-1)
+O(e^{-\kappa}).\label{x3}
\end{equation}
In both cases the integration on the right hand side should run
from $x=2$ to $x=4$, so extending the integrals to $\infty$ only
introduces errors of order $e^{-\kappa}$. In the OBM, $\tau$ is
determined by the condition that the first-order correction to the
free energy vanishes
\begin{equation}
\int_{2}^{\infty}\mbox{d}x\;x^{2}B(x)=-\frac{2\lambda}{3}+
G\int_{2}^{\infty}\mbox{d}x\;x^{2}(1+H(x-2))(e^{-U_{Y}(x)/k_{B}T}-1)
+O(e^{-\kappa})=0. \label{x2}
\end{equation}
This expression is used to rewrite Eqs. (\ref{x1}) and (\ref{x3})
\begin{equation}
\int_{2}^{\infty}\mbox{d}x\;xB(x)=-\frac{1}{2}
G\int_{2}^{\infty}\mbox{d}x\;x(x-2)(1+H(x-2))(e^{-U_{Y}(x)/k_{B}T}-1)
+O(e^{-\kappa}),
\end{equation}
\begin{equation}
\int_{2}^{\infty}\mbox{d}x\;x^{3}B(x)=
G\int_{2}^{\infty}\mbox{d}x\;x^{2}(x-2)(1+H(x-2))(e^{-U_{Y}(x)/k_{B}T}-1)
+O(e^{-\kappa})
\end{equation}
which are readily approximated. We substitute
$y=\exp[-\kappa(x/2-1)]$ which ultimately leads to
\begin{equation}
\int_{2}^{\infty}\mbox{d}x\;xB(x)=\frac{4G}{\kappa^{2}}J_{1}(\beta)+O(\kappa^{-3})
\label{x1b}
\end{equation}
and
\begin{equation}
\int_{2}^{\infty}\mbox{d}x\;x^{3}B(x)=
-\frac{16G}{\kappa^{2}}J_{1}(\beta)+O(\kappa^{-3}).\label{x3b}
\end{equation}
Here we have introduced
\begin{equation}
J_{1}(\beta)\equiv-\int_{0}^{1}\mbox{d}y\;\frac{e^{\beta
y}-1}{y}\ln y.
\end{equation}
An approximation for $J_{1}(\beta)$ that is accurate to within
1.4\% in the relevant range $0\leq\beta\leq 3.52$ is given by
\begin{equation}
J_{1}(\beta)\simeq\left\{
\begin{array}
[c]{ll}
\beta+\frac{1}{8}\beta^{2} & 0\leq\beta<0.8\\
2.81\left(e^{0.34\beta}-1\right)\hspace{10pt} & 0.8\leq \beta\leq
3.52.
\end{array}
\right.\label{J1}
\end{equation}
Finally, we insert Eqs. (\ref{x1b}) and (\ref{x3b}) into Eq.
(\ref{Y}). We thus obtain an approximation for the second-order
correction to the free energy
\begin{equation}
\Delta\simeq\frac{24G^{2}}{\kappa^{4}}\left(9G+10GH-18+\frac{\lambda}{2}\right)
J^{2}_{1}(\beta)\eta^{2}.\label{Delta2}
\end{equation}
In Table II we present typical values of $\Delta$. The corrections
to the pressure are very small (compare with Figs. 1 and 2).

\section*{Tables}

\hspace{0.3in}%
\begin{tabular}
[c]{|c|c|c|c|c|c|c|c|c|}

\hline & \multicolumn{4}{c|}{$\kappa=7$} & \multicolumn{4}{c|}{$\kappa=9$}\\

\hline \hspace{10pt}$\beta$\hspace{10pt} &
\hspace{15pt}$\eta_{l}$\hspace{15pt} &
\hspace{15pt}$\eta_{s}$\hspace{15pt} & \hspace{5pt}$\tau$
(OBM)\hspace{5pt}& \hspace{5pt}$\tau_{0}$ ($B_{2}$)\hspace{5pt} &
\hspace{15pt}$\eta_{l}$\hspace{15pt} &
\hspace{15pt}$\eta_{s}$\hspace{15pt} & \hspace{5pt}$\tau$
(OBM)\hspace{5pt}
& \hspace{5pt}$\tau_{0}$ ($B_{2}$)\hspace{5pt}\\

\hline $0$ & $0.492$ & $0.543$ & $\infty$ & $\infty$ & $0.492$ &
$0.543$ & $\infty$ & $\infty$\\

\hline $0.25$ & $0.493$ & $0.551$ & $8.921$ & $1.930$ & $0.493$ &
$0.552$ & $6.278$ & $2.549$\\

\hline $0.5$ & $0.493$ & $0.561$ & $3.329$ & $0.910$ & $0.494$ &
$0.563$ & $2.673$ & $1.199$\\

\hline $0.75$ & $0.492$ & $0.571$ & $1.689$ & $0.570$ & $0.494$ &
$0.576$ & $1.521$ & $0.750$\\

\hline $1$ & $0.490$ & $0.584$ & $0.986$ & $0.400$ & $0.492$ &
$0.591$ & $0.970$ & $0.526$\\

\hline $1.25$ & $0.485$ & $0.598$ & $0.616$ & $0.299$ & $0.488$ &
$0.608$ & $0.659$ & $0.392$\\

\hline $1.5$ & $0.478$ & $0.613$ & $0.405$ & $0.231$ & $0.480$ &
$0.626$ & $0.462$ & $0.303$\\

\hline $1.75$ & $0.465$ & $0.627$ & $0.271$ & $0.184$ & $0.464$ &
$0.643$ & $0.328$ & $0.240$\\

\hline $2$ & $0.441$ & $0.641$ & $0.183$ & $0.148$ & $0.437$ &
$0.657$ & $0.236$ & $0.193$\\

\hline
\end{tabular}
\smallskip

\vspace{5pt}

TABLE I: Volume fraction of particles in the coexisting fluid and
solid phases as a function of $\beta$ and $\kappa$ determined by
the simulations. The stickiness parameter is computed via the
Optimized Baxter Model (OBM) and the $B_{2}$ method ($B_{2}$).

\vspace{15pt}

\hspace{0.3in}%
\begin{tabular}
[c]{|c|c|c|c|c|c|c|c|}

\hline \hspace{10pt}$\tau$\hspace{10pt} &\hspace{10pt}$\kappa$\hspace{10pt} & $5$ & $7$ & $9$ & $11$ & $13$ & $15$\\

\hline \hspace{10pt}0.1\hspace{10pt} & $\beta$ &
\hspace{10pt}$1.966$\hspace{10pt} &
\hspace{10pt}$2.447$\hspace{10pt} &
\hspace{10pt}$2.805$\hspace{10pt} &
\hspace{10pt}$3.088$\hspace{10pt} &
\hspace{10pt}$3.321$\hspace{10pt} &
\hspace{10pt}$3.519\hspace{10pt}$\\

\hline & $\Delta$ & $-0.1315$ & $-0.0637$ & $-0.0352$ & $-0.0214$
& $-0.0139$ & $-0.0096$\\

\hline 0.15 & $\beta$ & $1.725$ & $2.099$ & $2.397$ &
$2.644$ & $2.855$ & $3.038$\\

\hline & $\Delta$ & $-0.2129$ & $-0.0944$ & $-0.0505$ & $-0.0303$
& $-0.0197$ & $-0.0135$\\

\hline 0.2 & $\beta$ & $1.548$ & $1.853$ & $2.110$ &
$2.332$ & $2.526$ & $2.697$\\

\hline & $\Delta$ & $-0.2762$ & $-0.1154$ & $-0.0603$ & $-0.0360$
& $-0.0233$ & $-0.0160$\\

\hline 0.5 & $\beta$ & $1.017$ & $1.140$ & $1.282$ &
$1.423$ & $1.558$ & $1.686$\\

\hline & $\Delta$ & $-0.4291$ & $-0.1467$ & $-0.0715$ & $-0.0416$
& $-0.0268$ & $-0.0186$\\

\hline 1 & $\beta$ & $0.694$ & $0.727$ & $0.803$ &
$0.891$ & $0.981$ & $1.070$\\

\hline & $\Delta$ & $-0.4209$ & $-0.1211$ & $-0.0549$ & $-0.0312$
& $-0.0200$ & $-0.0139$\\

\hline
\end{tabular}
\smallskip

\vspace{5pt}

TABLE II: Estimates of the second-order correction $\Delta$ to the
free energy for $\eta=0.4$ at various $\tau$ and $\kappa$.
$\Delta$ is determined from Eq. (\ref{Delta2}) within the
approximation given by Eq. (\ref{J1}).

\section*{Figure Captions and Figures}

FIG. 1. Dimensionless chemical potential and dimensionless
pressure as a function of the strength $\beta$ of the Yukawa
potential for the coexisting fluid and solid phases. Here
$\kappa=7$. The diamonds and the fitted line are results from the
simulations. The squares are predictions from the Optimized Baxter
Model (at the same densities) and the triangles have been computed
by the $B_{2}$ method.

FIG. 2. Same as Fig. 1 but now at $\kappa=9$.

FIG. 3. Dimensionless chemical potential and dimensionless
pressure as a function of the inverse range $\kappa$ of the Yukawa
potential at volume fraction $\eta=0.1$. The solid symbols are
results from the simulations, the horizontal lines are predictions
from the Optimized Baxter Model at a variety of fixed values of
$\tau$. In the simulations the strength of the attraction $\beta$
is chosen in such a way that the Optimized Baxter Model gives the
appropriate value of $\tau$: grey filled diamonds $\tau=1$, grey
filled squares $\tau=0.5$, black filled triangles $\tau=0.2$,
black filled squares $\tau=0.15$ and black filled diamonds
$\tau=0.1$. The corresponding open symbols have been computed by
the $B_{2}$ method.

FIG. 4. Same as Fig. 3 but now at volume fraction $\eta=0.2$.

FIG. 5. Same as Fig. 3 but now at volume fraction $\eta=0.3$.

FIG. 6. Same as Fig. 3 but now at volume fraction $\eta=0.4$.

FIG. 7. Example of the dependence of the average dimensionless
energy per particle $\langle U_{1}/k_{B}T\rangle_{N}$ in the fluid
on the size of the system. Here $\kappa=9$, $\beta=1$ and
$\eta=\pi/15\approx 0.20944$ ($\rho (2a)^{3}=0.4$). $N$ denotes
the number of particles.

FIG. 8. Same as Fig. 7 but now for the solid at $\eta=\pi/5\approx
0.62832$ ($\rho (2a)^{3}=1.2$).

\hspace{60pt}

  \begin{figure}[h]
  \begin{minipage}[t]{.45\textwidth}
    \begin{center}
      \epsfig{file=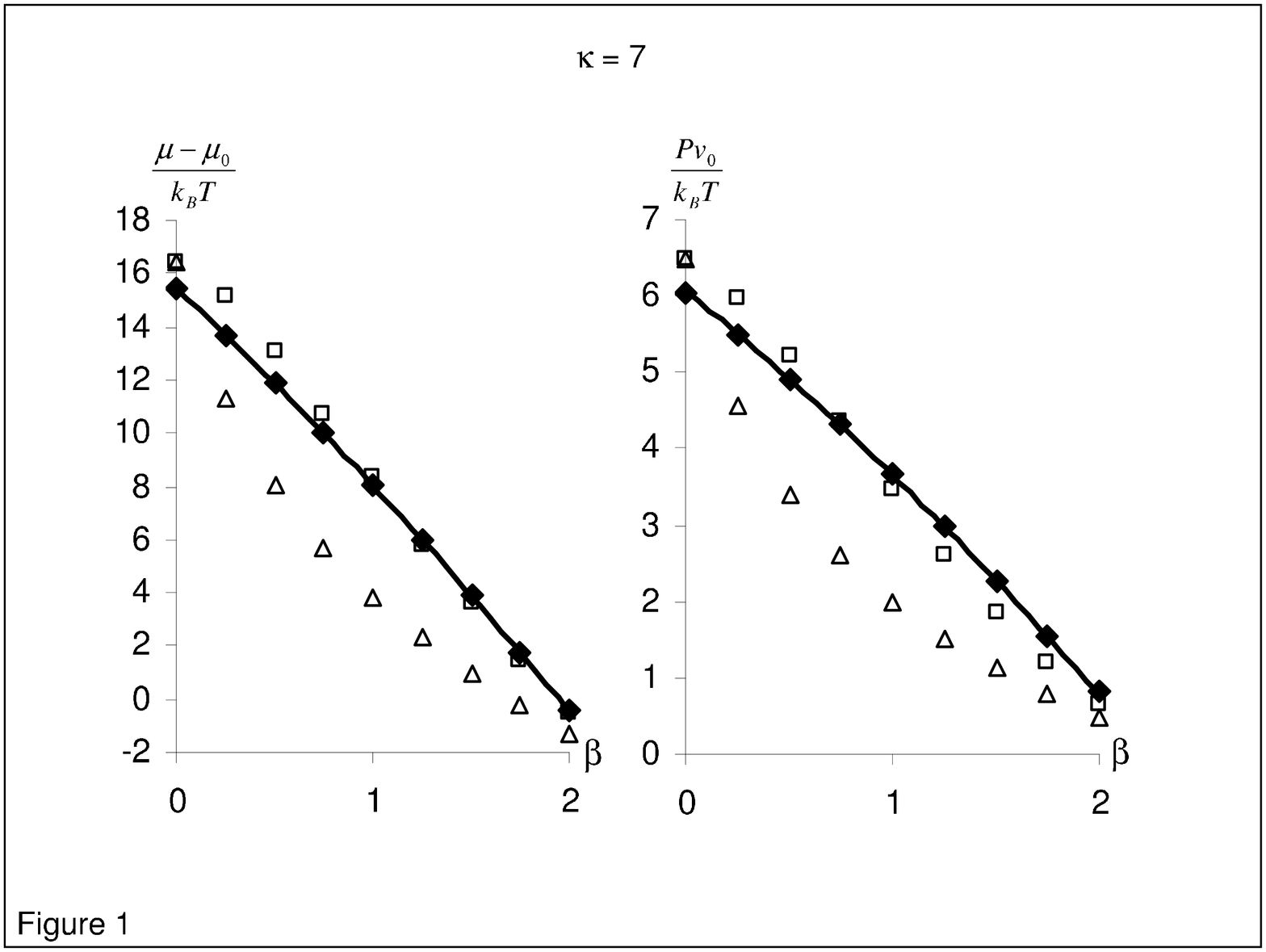, scale=0.5, angle=270, width=250pt}
    \end{center}
  \end{minipage}
  \begin{minipage}[t]{.45\textwidth}
    \begin{center}
      \epsfig{file=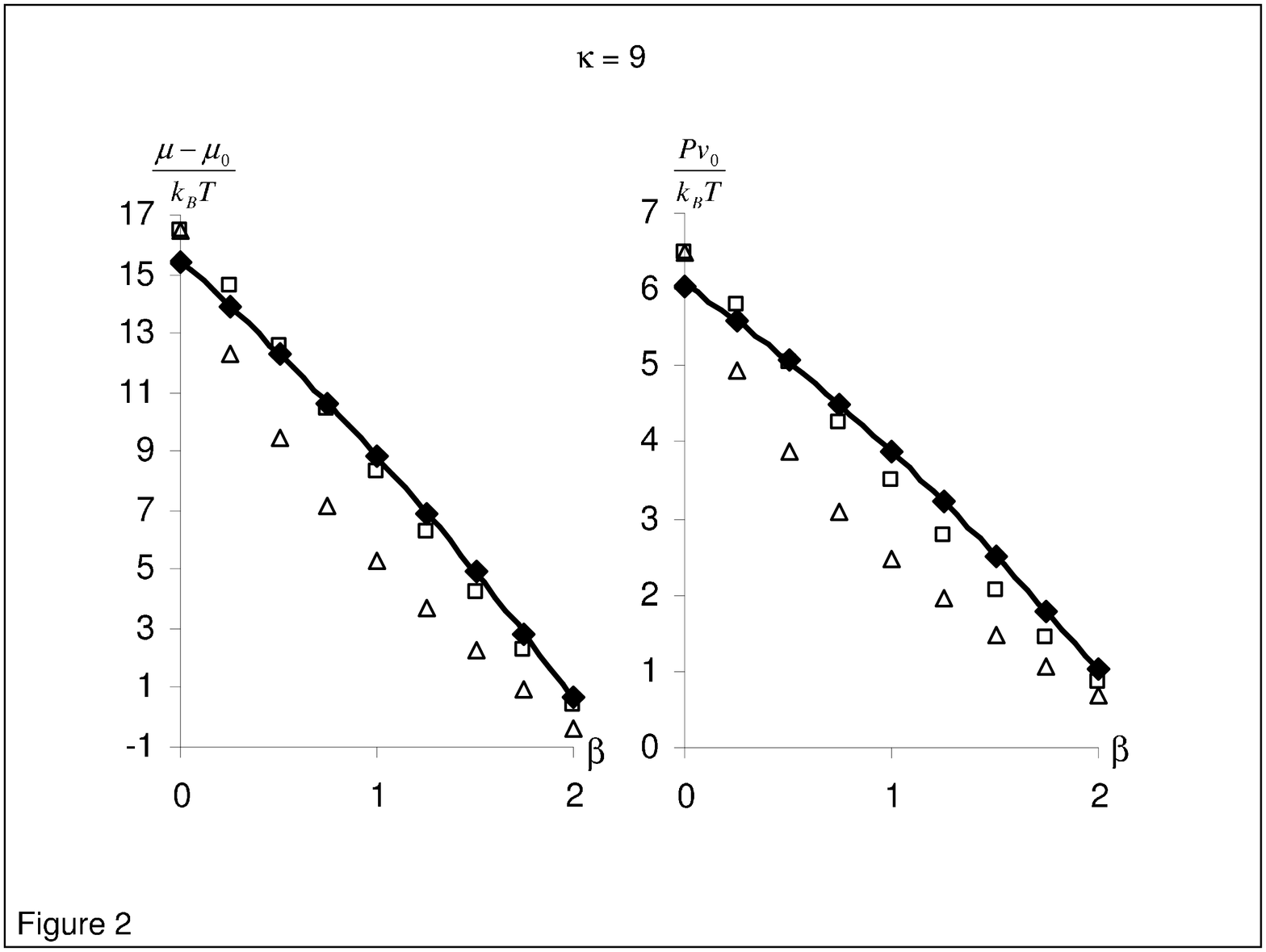, scale=0.5, angle=270, width=250pt}
    \end{center}
  \end{minipage}
  \end{figure}
  \begin{figure}[h]
  \begin{minipage}[t]{.45\textwidth}
    \begin{center}
      \epsfig{file=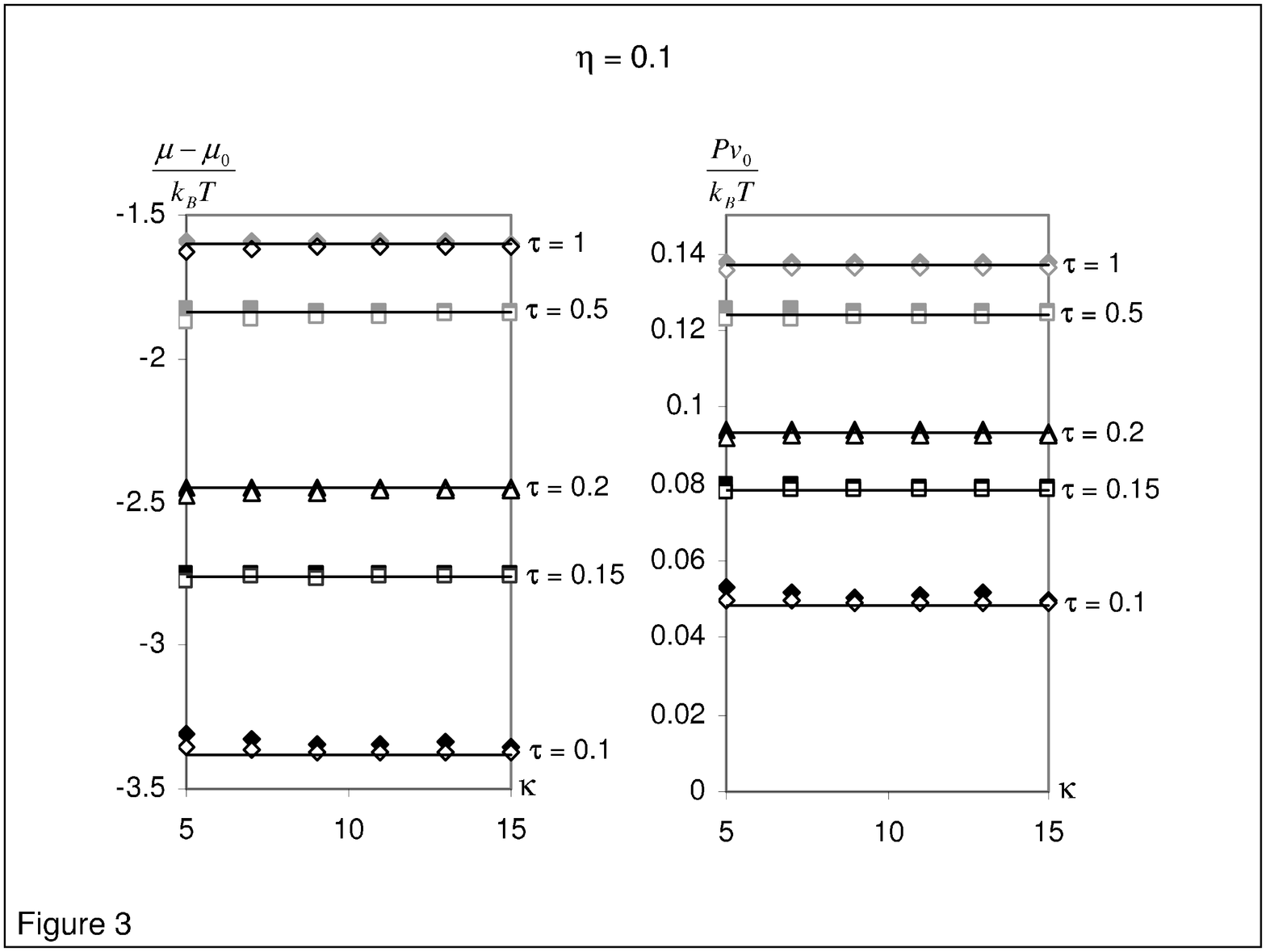, scale=0.5, angle=270, width=250pt}
    \end{center}
  \end{minipage}
  \begin{minipage}[t]{.45\textwidth}
    \begin{center}
      \epsfig{file=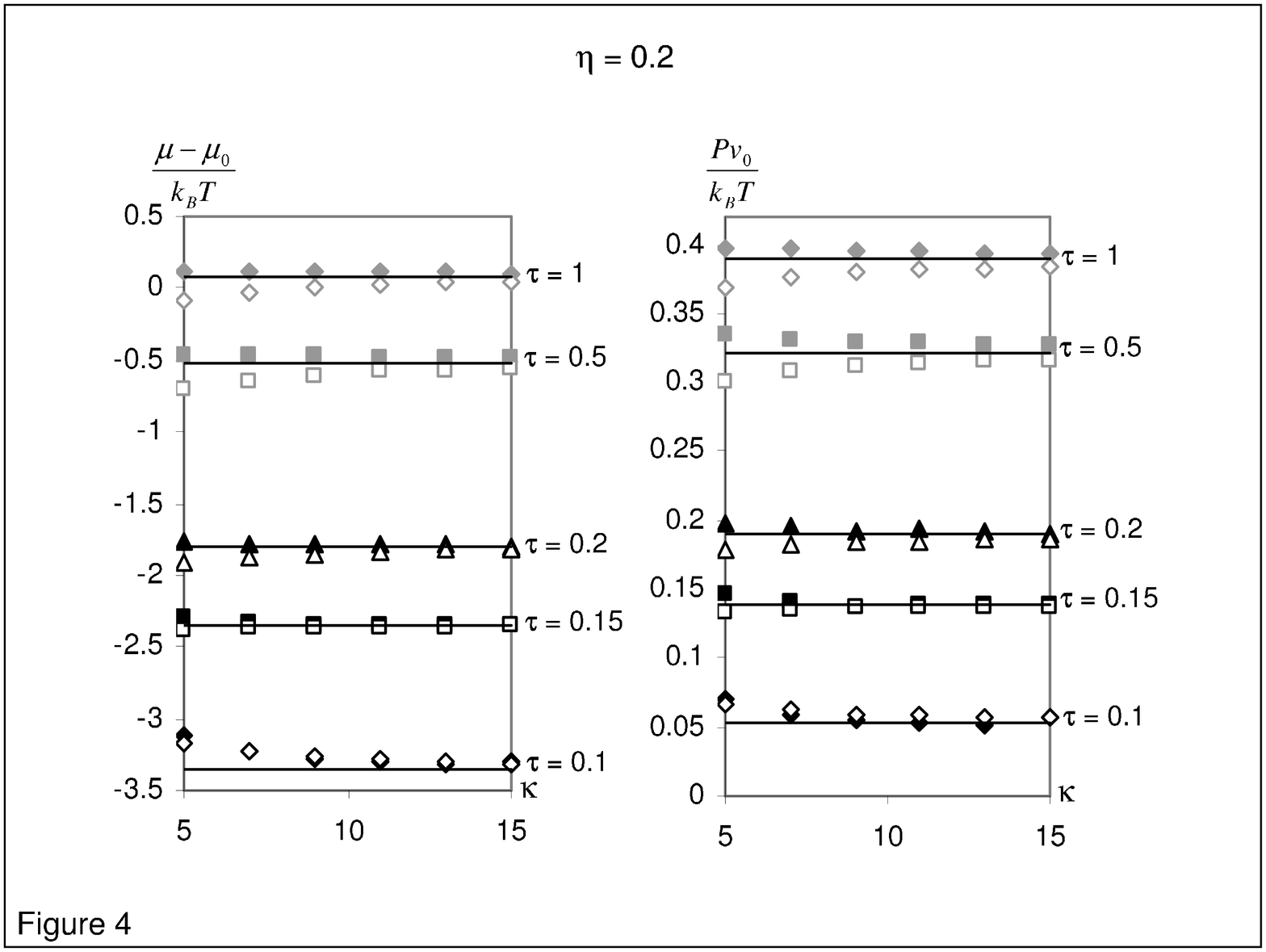, scale=0.5, angle=270, width=250pt}
    \end{center}
  \end{minipage}
  \begin{minipage}[t]{.45\textwidth}
    \begin{center}
      \epsfig{file=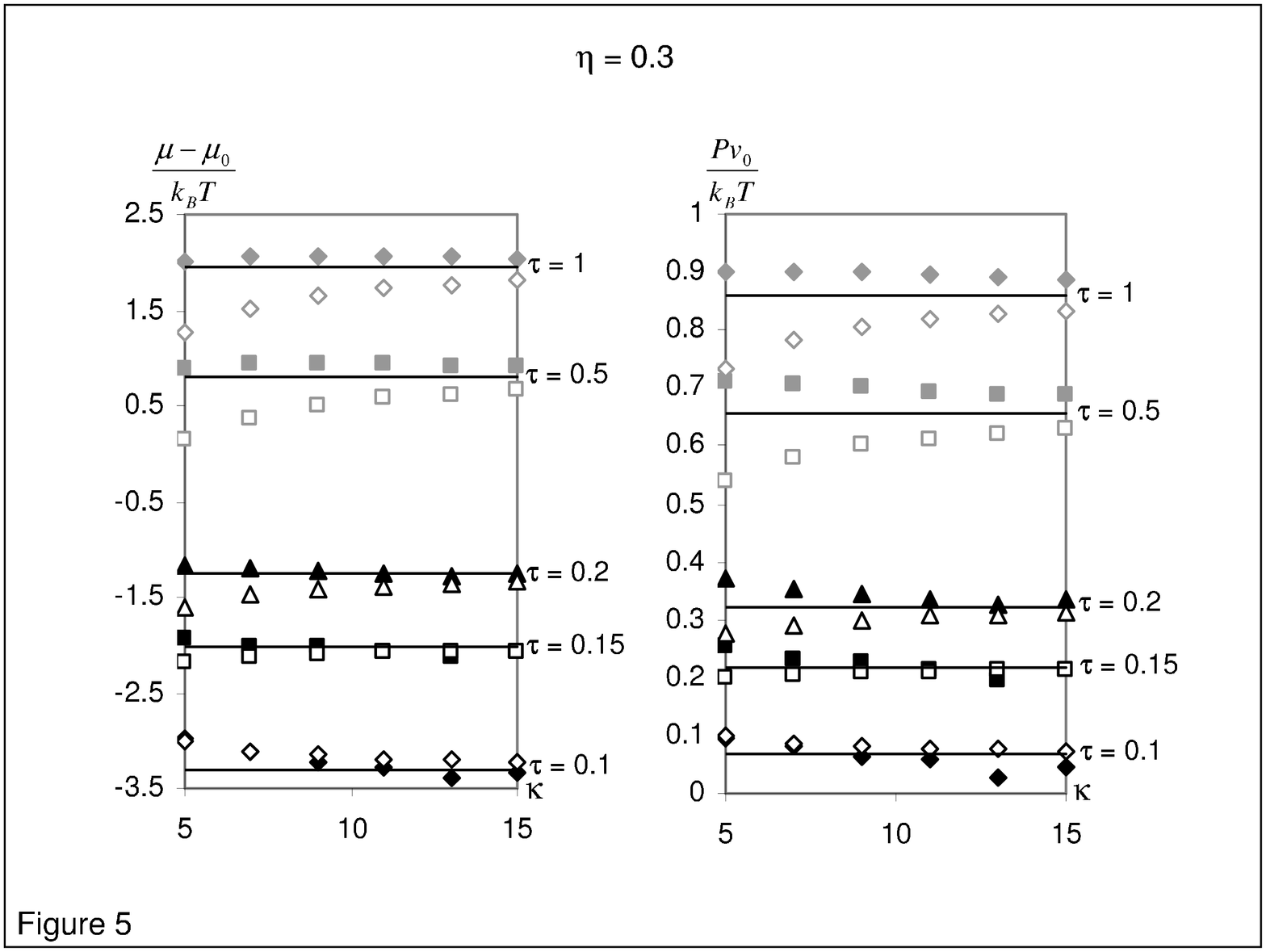, scale=0.5, angle=270, width=250pt}
    \end{center}
  \end{minipage}
  \begin{minipage}[t]{.45\textwidth}
    \begin{center}
      \epsfig{file=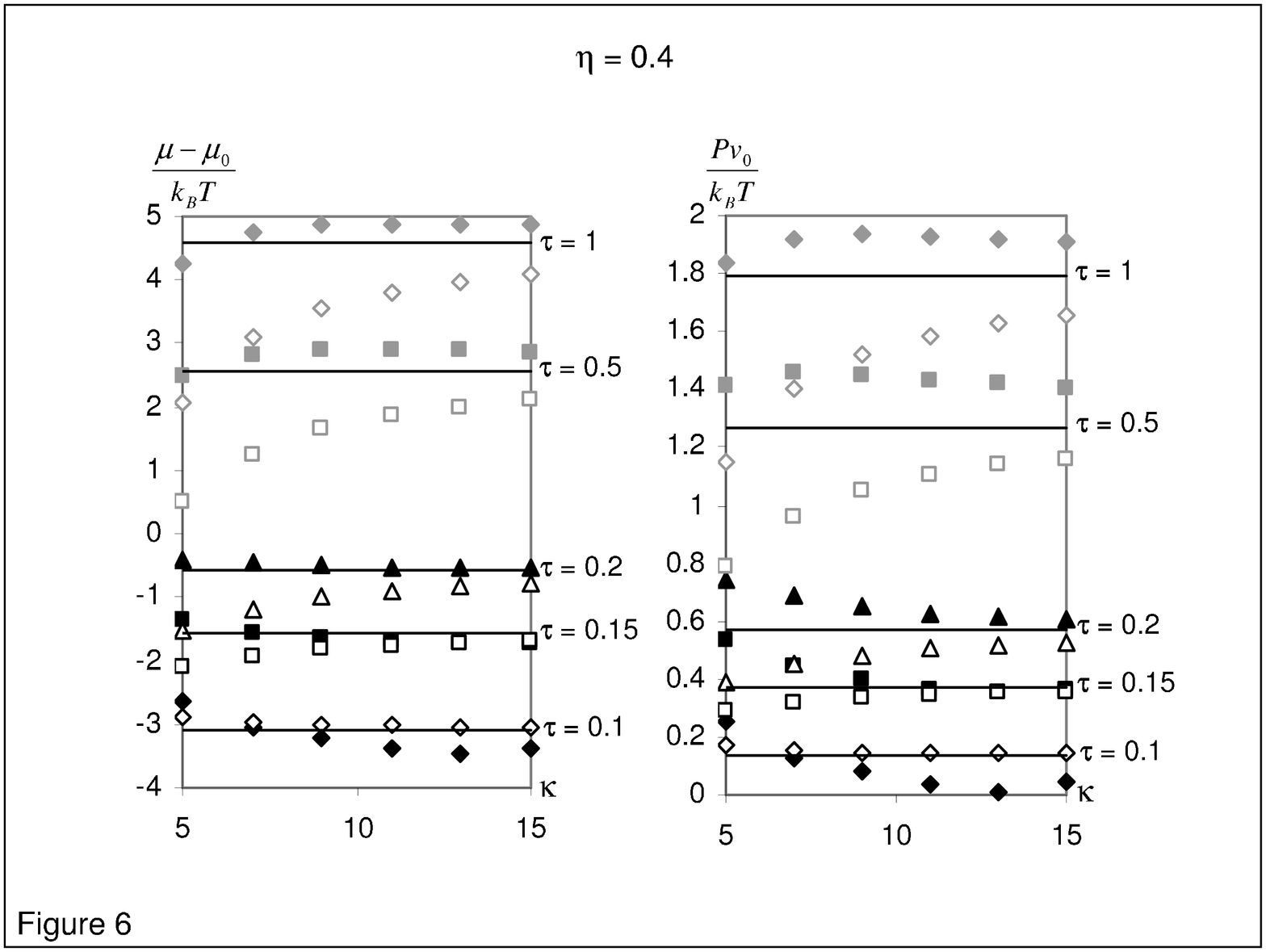, scale=0.5, angle=270, width=250pt}
    \end{center}
  \end{minipage}
  \begin{minipage}[t]{.45\textwidth}
    \begin{center}
      \epsfig{file=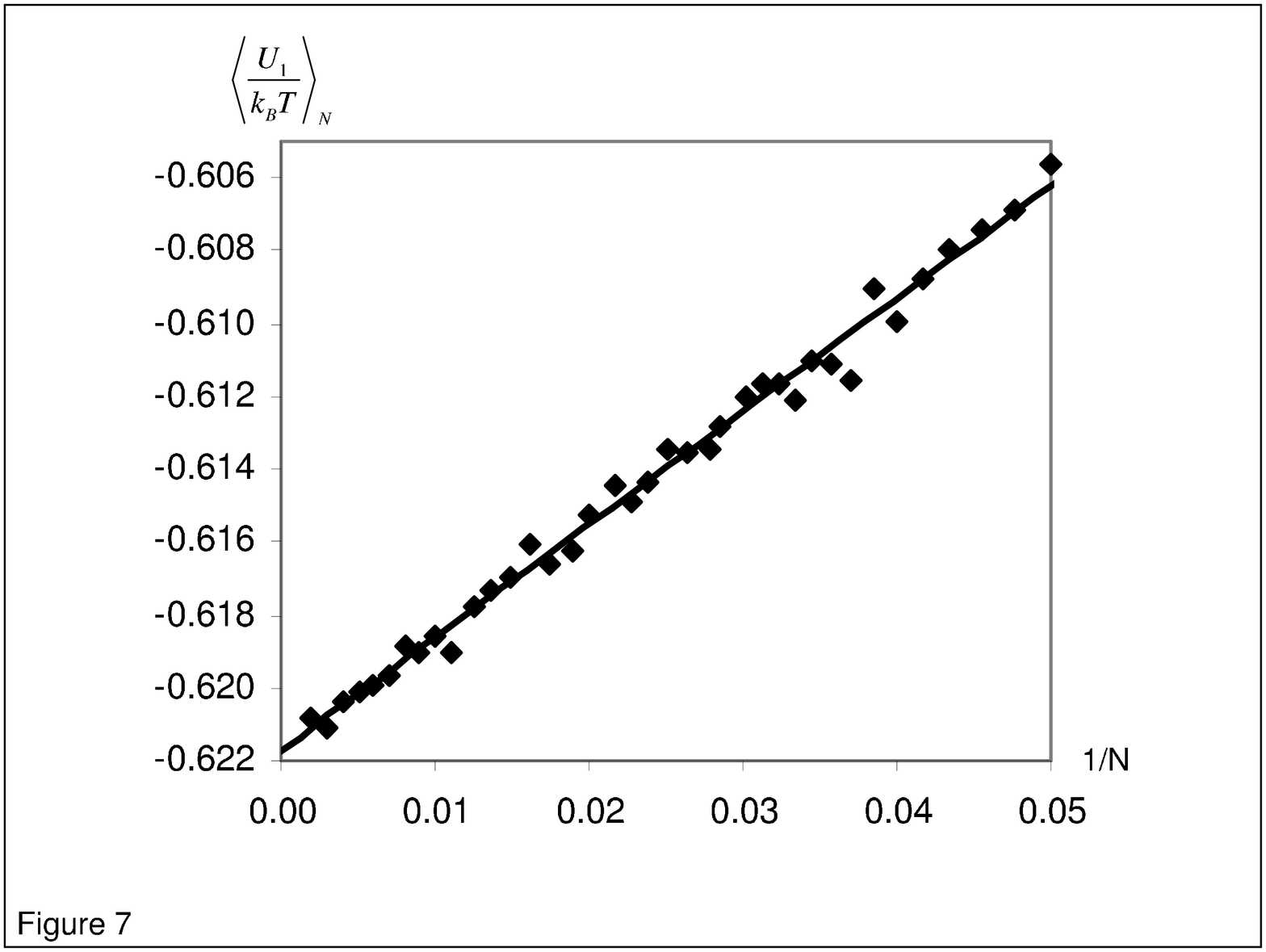, scale=0.5, angle=270, width=240pt}
    \end{center}
  \end{minipage}
  \begin{minipage}[t]{.45\textwidth}
    \begin{center}
      \epsfig{file=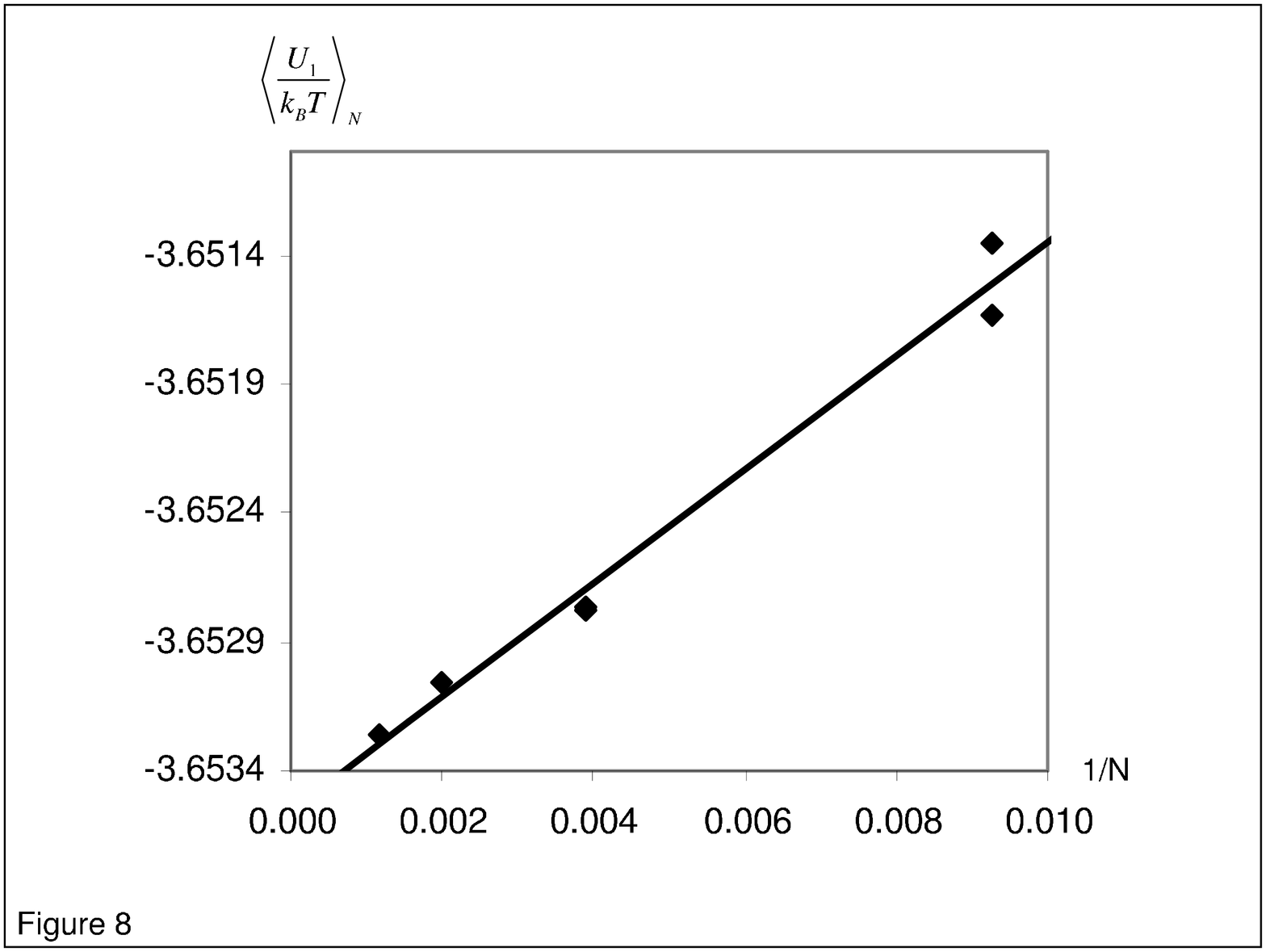, scale=0.5, angle=270, width=240pt}
    \end{center}
  \end{minipage}
  \end{figure}

\end{document}